\documentclass[a4paper,10pt]{article}
\usepackage[xdvi]{graphics}
\usepackage{amssymb}
\usepackage{latexsym}
\usepackage{amsmath}
\usepackage{epsf}
\newcommand{\e}{{\rm e}}

\newcommand{\EP}{\mathcal{E}}

\def\eqalign#1{\,\vcenter{\openup1\jot\ialign
    {\strut\hfil$\displaystyle{##}$&$\displaystyle{{}##}$
      \hfil\crcr#1\crcr}}\,}
\def\aoverchi{1-\frac{a}{\chi}}
\numberwithin{equation}{section}
\title{Comment on the exterior solutions and their geometry
in scalar-tensor theories of gravity}
\author{
Tooru TSUCHIDA  
\thanks{Electronic address:tsuchida@astro2.sc.niigata-u.ac.jp} \,  
and Kazuya WATANABE
\thanks{Electronic address:kazuya@astro2.sc.niigata-u.ac.jp}\\ 
{\em Department of Physics,~Niigata University, Niigata 950-2181, Japan.}}
\date{}
\begin{document}
\maketitle


\begin{abstract}

We study series of the stationary solutions
with asymptotic flatness properties 
in the Einstein-Maxwell-free scalar system 
because they are locally equivalent with 
the exterior solutions in some class 
of the scalar-tensor theories of gravity.
First, we classify spherical exterior solutions
into two types of the solutions, an apparently black hole type solution
and an apparently worm hole type solution.
The solutions contain three parameters, and
we clarify their physical significance.
Second, we reduce the field equations 
for the axisymmetric exterior solutions.
We find that the reduced equations are partially the same as 
the Ernst equations.
As simple examples, we derive new series of the static, axisymmetric 
exterior solutions, which correspond to Voorhees's solutions.
We then show a non-trivial relation between
the spherical exterior solutions and our new solutions. 
Finally, since null geodesics have conformally invariant properties,
we study the local geometry of the exterior solutions
by using the optical scalar equations
and find some anomalous behaviors of the null geodesics.
\end{abstract}

\

\section{Introduction}

\

Recently, as natural alternatives to general relativity, 
the scalar-tensor theories have been studied by many theoretical
physicists. In these theories, the gravity is mediated not only by a
tensor field but also by a scalar field. 
Also, such theories have been of interest
as effective theories of the string theory at low energy scales \cite{1}.

Several theoretical predictions in the scalar-tensor theories
have been obtained (see e.g. Ref.\cite{2}$\sim$\cite{55}). 
It has been found that a wide class of the scalar-tensor theories can pass 
all the experimental tests in weak gravitational fields.
However, it has also been found that the scalar-tensor theories
show different aspects of the gravity in the
strong gravitational fields in contrast 
to general relativity (see e.g. Ref.\cite{55}).
It has been shown numerically 
that nonperturbative effects in the scalar-tensor 
theories increase the maximum mass of an isolated system
such as a neutron star [2-5].  
In these works, numerical methods play an important role,
and, in interpreting the numerical results,
a static, spherically symmetric vacuum solution
(a spherical exterior solution) is matched to
the numerical solution.

The spherical exterior solution is an analytic exact solution,
and, as far as the authors know, only few exact solutions
have been known. In particular,
the axisymmetric exterior solution corresponding to
the Kerr solution must be of significant interest.  
Motivated by this fact, 
we study series of the stationary solutions
with asymptotic flatness properties 
in the Einstein-Maxwell-free scalar system 
because, as summarized
in Appendix A,
they are locally equivalent with 
the exterior solutions in some class 
of the scalar-tensor theories of gravity.
The field equations are given by
\begin{equation} 
\eqalign{
&
R_{\mu\nu}=8\pi T_{\mu\nu}+2\varphi_\mu\varphi_\nu,
\cr
&
\Box\varphi=0,
\cr
}
\end{equation}
where $T_{\mu\nu}$ is a energy-momentum tensor of 
the Maxwell field (see also Appendix A). 

In this paper we first classify spherical exterior solutions
into two types of the solutions, an apparently black hole type (ABH)
solution and an apparently worm hole type (AWH) solution.
We then clarify physical significance of the parameters
contained in the solutions and show that
the ABH solution and the AWH solution correspond to, respectively, 
the mass dominant case and the charge dominant case
of the Reissner-Nordstr\"om solution.
Second, we reduce the field equations 
for the axisymmetric exterior solutions
and show that the reduced equations are partially the same as 
the Ernst equations.
As simple examples, we derive new series of static, axisymmetric 
exterior solutions, which will be referred to as 
scalar-tensor-Weyl solutions.
We then show a non-trivial relation between
the spherical exterior solutions and our new solutions. 
Finally, we study the local geometry of the exterior solutions
by using the optical scalar equations
and find some anomalous behaviors of the null geodesics.

\

\section{Spherical exterior solutions}

\subsection{Classification of the exterior solutions}

\

In this section, we derive static, spherically 
symmetric exterior solutions with the electric field,
which are hereafter referred to as spherical exterior solutions. 
In the Just coordinate, the metric becomes \cite{4}:  
\begin{equation}
ds^2=-\e^{\gamma}dt^2+\e^{-\gamma}d\chi^2+\e^{\lambda-\gamma}d\Omega^2,
\label{e1}
\end{equation}
where $\gamma$ and $\lambda$ are functions of $\chi$.
Non-vanishing components of the electromagnetic tensor,
$F_{\mu\nu}$, are $F_{01}=-F_{10}=E(\chi)$.
The Maxwell equation is reduced to
\begin{equation}
(E\e^{\lambda-\gamma})'=0,
\label{e3}
\end{equation}
where a prime denotes a derivative with respect to $\chi$.
With (\ref{e3}) one obtains 
\begin{equation}
E=Q\e^{\gamma-\lambda},
\label{e4}
\end{equation}
where the integration constant, $Q$, is an electric charge.

The Einstein equation and the field equation for $\varphi$ become
\begin{subequations}
\begin{eqnarray}
\frac{1}{2}(\gamma''+\lambda'\gamma')\e^{\gamma} & = &
Q^2\e^{2(\gamma-\lambda)}, 
\label{e5} \\
\frac{1}{2}(\gamma''-2\lambda''-{\lambda'}^{2}-{\gamma'}^{2}
+\lambda'\gamma')\e^{\gamma}
& = & 2{\varphi'}^{2}\e^{\gamma}-Q^2\e^{2(\gamma-\lambda)}, 
\label{e6} \\
\e^{\gamma-\lambda}
+\frac{1}{2}(\gamma''-\lambda''+\lambda'\gamma'-{\lambda'}^{2})\e^{\gamma}
& = & Q^2\e^{2(\gamma-\lambda)}, 
\label{e7} \\
(\varphi''+\lambda'\varphi')\e^{\gamma}&=&0.
\label{e8}
\end{eqnarray}
\end{subequations}
With (\ref{e5}) and (\ref{e7}), one obtains
\begin{equation}
(\e^{\lambda})'' = 2.
\label{e9}
\end{equation}
One finds two types of the solutions to (\ref{e9}), which are
referred to as
an {\it apparently black hole type} (ABH) solution and 
an {\it apparently worm hole type} (AWH) solution, respectively: 
\begin{subequations}
\begin{eqnarray}
\mbox{ABH}\hspace{1cm} 
\e^{\lambda}& =&\chi^2-a\chi \hspace{1cm} (\chi > a \ge 0) , 
\label{e10} \\
\mbox{AWH}\hspace{1cm} 
\e^{\lambda}&=& \chi^2+\frac{a^2}{4} \hspace{1cm} (-\infty < \chi
<\infty ) ,
\label{e11}
\end{eqnarray}
\end{subequations}
where $a$ is an integration constant.

When $a\neq0$, the equation (\ref{e8}) is integrated as
\begin{subequations}
\begin{eqnarray}
\mbox{ABH}\hspace{1cm}
\varphi & = & \varphi_0 + \frac{c}{a}\ln\left(1-\frac{a}{\chi}\right), 
\label{e13} \\
\mbox{AWH}\hspace{1cm}
\varphi & = & \varphi_0 +
\frac{2c}{a}\left[\arctan\left(\frac{2\chi}{a}\right)-\frac{\pi}{2}\right],
\label{e14}
\end{eqnarray}
\end{subequations}
where $\varphi_0$ and $c$ are
integration constants [2-5].

The equation $(\ref{e5})$ is integrated as
\begin{equation}
(\gamma'\e^{\lambda})^2 =4Q^2\e^{\gamma}+\epsilon b^2 ,
\label{e16}
\end{equation}
where $b$ is an integration constant, and 
$\epsilon$ is 1 in the ABH solution and 
$-1$ in the AWH solution.

With (\ref{e5}), (\ref{e6}) and (\ref{e7}), 
we obtain 
\begin{equation}
\frac{{\lambda'}^2-{\gamma'}^2}{4}-\e^{-\lambda}=
{\varphi'}^2-Q^2\e^{\gamma-2\lambda},
\label{e17}
\end{equation}
which is reduced to
the following relation among the integration constants:
\begin{subequations}
\begin{eqnarray}
\mbox{ABH}\hspace{1cm}
a^2-b^2 & = & 4c^2 \longrightarrow a\ge b \ge 0 ,
\label{e18} \\
\mbox{AWH}\hspace{1cm}
b^2-a^2 & = & 4c^2 \longrightarrow b\ge a \ge 0 .
\label{e19}
\end{eqnarray}
\end{subequations}
Since $\e^{\gamma}\longrightarrow 1$ as $\chi\longrightarrow\infty$
in the asymptotically flat spacetime, with (\ref{e16}),
we obtain the following inequality for the AWH solution:
\begin{equation}
b^2 \le 4Q^2.
\label{e20}
\end{equation}
That is,
the AWH solution is {\it charge dominant}, and 
no AWH solution exists when $Q=0$.

\
 
\subsection{The ABH solution}

\

The metric function, $\gamma$, in the ABH solution is
obtained as follows:
\begin{subequations}
\begin{equation}
\e^{\gamma} = \frac{B}{(1-B)^2}\left(\frac{b}{Q}\right)^2 ,
\label{e26}
\end{equation}
where 
\begin{equation}
\begin{array}{cc}
B \equiv k\left(1-\dfrac{a}{\chi}\right)^{\frac{b}{a}},&
k \equiv \dfrac{\sqrt{4Q^2+b^2}-b}{\sqrt{4Q^2+b^2}+b}\ <1.
\end{array}
\label{e28}
\end{equation}
\end{subequations}
In the limit, $Q\rightarrow 0$,
the solution coincides with 
the previously known solution [2-5].
In the limit, $b\rightarrow 0$, 
the solution is reduced to the following form:
\begin{equation}
\e^{\gamma}
=\left[1-\frac{Q}{a}\ln\left(1-\frac{a}{\chi}\right)\right]^{-2}.
\label{e21}
\end{equation}

When $a=b$, i.e., $c=0$,
the Just coordinate, $\chi$, is related to the Schwarzschild
coordinate, $r$, as
\begin{equation}
\chi = r-\frac{kb}{1-k} .
\label{e35}
\end{equation}
The metric function, $\gamma$, becomes 
\begin{subequations}
\begin{equation}
\e^{\gamma} = \left[1-\left(\frac{kb}{1-k}\right)\frac{1}{r}\right]
\left[1-\left(\frac{b}{1-k}\right)\frac{1}{r}\right]
\equiv \left(1-\frac{r_{-}}{r}\right)\left(1-\frac{r_{+}}{r}\right) 
\equiv 1-\frac{2m}{r}+\frac{Q^2}{r^2} ,
\label{e37}
\end{equation}
where
\begin{equation}
r_{+} \equiv \frac{b}{1-k} ,
\ \
r_{-} \equiv \frac{kb}{1-k} ,
\ \
2m    \equiv \frac{1+k}{1-k}b = \sqrt{4Q^2+b^2} .
\label{e41}
\end{equation}
\end{subequations}
With (\ref{e41}), we obtain the following inequality:
\begin{equation}
0 \le b^2=4(m^2-Q^2).
\label{e42}
\end{equation}
That is, the ABH solution corresponds to the {\it mass dominant}
($m^2>Q^2$) Reissner-Nordstr\"om spacetime
and has two event horizons, $r_+$ and $r_-$, when $a=b$.

When $a>b$,
the Just coordinate, $\chi$, is related to the Schwarzschild
coordinate, $r$, as
\begin{equation}
r^2 =  \chi^2\left(1-\frac{a}{\chi}\right)
\frac{(1-B)^2}{B}\left(\frac{Q}{b}\right)^2, 
\end{equation}
which vanishes at 
$\chi=a$, $0$ and 
$\chi_B\equiv{a}/\left({1-k^{-a/b}}\right)<0$.
Moreover, we find that 
$\chi=a$, $\chi=0$ and $\chi=\chi_B$,
correspond to, respectively, 
$r=r_{+}$, $r=r_{-}$ and $r=0$ when $a=b$ and that
the null surface, $\chi=a$, becomes a singularity when $a>b$.  

\

\subsection{The AWH solution}

\

When $a>0$, the metric function, $\gamma$, in the AWH solution
is obtained as
\begin{subequations}
\begin{equation}
\e^{\gamma}=\frac{1}{4}\left(\frac{b}{Q}\right)^2\sec^2\left
[\frac{b}{a}\arctan\left(\frac{2\chi}{a}\right)+\beta\right], 
\label{e29}
\end{equation}
where the constant, $\beta$, is defined by 
\begin{equation}
\sec^2\left[\frac{\pi b}{2 a}+\beta\right]=\frac{4Q^2}{b^2}.
\label{e30}
\end{equation}
\end{subequations}
As mentioned previously, the AWH solution does not exist when $Q=0$.
    
When $a=0$, the AWH solution becomes
\begin{subequations}
\begin{eqnarray}
\e^{\lambda} & = & \chi^2 ,
\label{e22} \\
\varphi & = & \varphi_0 -\frac{c}{\chi},
\label{e23} \\
\e^{\gamma} & = & \frac{c^2}{Q^2}\sec^2 \left(\beta-\frac{c}{\chi}\right),
\label{e24} \\
\end{eqnarray}
\end{subequations}
where
\begin{equation}
\sec^2\beta=
\frac{Q^2}{c^2}.
\label{e25}
\end{equation}

When $a=b$, the AWH solution is reduced to the following:  
\begin{subequations}
\begin{eqnarray}
\e^{\lambda} & = & \chi^2+\frac{b^2}{4}, 
\label{e44} \\
\e^{\gamma} & = & \frac{1}{4}\left(\frac{b}{Q}\right)^2
\sec^2\left[\arctan\left(\frac{2\chi}{b}\right)+\beta\right],
\label{e46} \\
\varphi & = & \varphi_0,
\label{e45} 
\end{eqnarray}
\end{subequations}
where 
\begin{equation}
\sec^2\left(\frac{\pi}{2}+\beta\right) =\frac{4Q^2}{b^2}.
\label{e47}
\end{equation}
Note that (\ref{e47}) is equivalent with the following:
\begin{equation}
\sin\beta  =  \pm \frac{b}{2Q}, \ \ 
\cos\beta  =  \sqrt{1-\frac{b^2}{4Q^2}} .
\label{e48}
\end{equation}
Then (\ref{e46}) becomes
\begin{equation}
\e^{\gamma} =
\frac{b^2}{4Q^2}\left(1+4\frac{\chi^2}{b^2}\right)
\left(\sqrt{1-\frac{b^2}{4Q^2}}\mp \frac{\chi}{Q}\right)^{-2},
\label{e49}
\end{equation}
and the Just coordinate, $\chi$, is related to the Schwarzschild
coordinate, $r$, as
\begin{equation}
r = \e^{(\lambda-\gamma)/2}=\chi\mp \sqrt{Q^2-\frac{b^2}{4}}
\equiv r^{\mp}.
\label{e50}
\end{equation}
When we take $r^-$, we have
\begin{equation}
\e^{\gamma} = 1+2\sqrt{Q^2-\frac{b^2}{4}}\frac{1}{r}+\frac{Q^2}{r^2}
\equiv 1-\frac{2m}{r}+\frac{Q^2}{r^2},
~~m\equiv -\sqrt{Q^2-\frac{b^2}{4}}<0,
\label{e51}
\end{equation}
which is a negative mass solution and should be discarded
as an unphysical exterior solution.
When we take $r^+$, we have
\begin{equation}
\e^{\gamma}= 1-2\sqrt{Q^2-\frac{b^2}{4}}\frac{1}{r}+\frac{Q^2}{r^2}
\equiv 1-\frac{2m}{r}+\frac{Q^2}{r^2},
~~0<m\equiv \sqrt{Q^2-\frac{b^2}{4}}.
\label{e52}
\end{equation}
That is, the AWH solution corresponds to 
the {\it charge dominant}
($m^2<Q^2$) Reissner-Nordstr\"om spacetime
and has a naked timelike singularity, $r=0$, when $a=b$.

When $a<b$,
the Just coordinate, $\chi$, is related to the Schwarzschild
coordinate, $r$, as
\begin{equation}
r^2  =  4\left(\frac{Q}{b}\right)^2
\left(\chi^2+\frac{a^2}{4}\right) 
\cos^2\left[\frac{b}{a}\arctan\left(\frac{2\chi}{a}\right)+\beta\right].
\label{e61}
\end{equation}
Note that $r=0$ 
at $\chi=\chi_{\mbox{\tiny W}}$, where 
\begin{equation}
\chi_{\mbox{\tiny W}}=
\frac{a}{2}\tan\left[\frac{a}{b}\left(\frac{\pi}{2}-\beta\right)\right].
\end{equation}
The AWH solution has a 
naked timelike singularity at $\chi=\chi_{\mbox{\tiny W}}$.
That is, the AWH solution does not have a worm hole,
and the asymptotic region defined 
by $\chi\rightarrow-\infty$ does not exist.

\

\section{The Ernst equations in scalar-tensor theories}

\

In the stationary, axisymmetric spacetime, 
there exist two Killing vectors, $\boldsymbol{\xi}=\partial_t$ and 
$\boldsymbol{\eta}=\partial_{\phi}$. 
It is shown that the energy-momentum tensor,
$T^{\alpha\beta}$, of the Maxwell field satisfies 
the following relations \cite{5}:
\begin{equation}
\xi^{\alpha}T_{\alpha}^{\ [\beta}\xi^{\gamma}\eta^{\delta]}=
\eta^{\alpha}T_{\alpha}^{\ [\beta}\xi^{\gamma}\eta^{\delta]}=
0.
\label{a01}
\end{equation}
Moreover, one finds that
the similar relations hold
for $T_{\mu\nu}^{[\varphi]}\equiv
2\partial_{\mu}\varphi\partial_{\nu}\varphi$. 
Accordingly, with the theorem 7.1.1 in Ref.\cite{20}, 
the metric can be reduced to the following form:
\begin{equation}
ds^2=-\e^{2\psi}(dt-\omega d\phi)^2 +
\e^{-2\psi}\left[\e^{2\gamma}(d\rho^2+dz^2)+\rho^2d\phi^2\right],
\label{a1}
\end{equation}
where the metric functions, $\psi,\gamma$ and $\omega$, 
are functions of $x^1\equiv\rho$ and $x^2\equiv z$.

After long and complicated calculations, we find that 
the field equations are reduced to
the following: 
\begin{subequations}
\begin{eqnarray}
\e^{2\psi} & = & \frac{1}{2}\left({\EP}+\bar{\EP}\right)+\Phi\bar{\Phi}, 
\label{a7} \\
\e^{2\psi}\nabla^2{\EP} & = &
(\nabla\EP)\cdot\left[(\nabla\EP)+2\bar{\Phi}\nabla\Phi\right], 
\label{a8} \\
e^{2\psi}\nabla^2{\Phi} & = &
(\nabla\Phi)\cdot\left[(\nabla\EP)+2\bar{\Phi}\nabla\Phi\right], 
\label{a9} \\
\gamma_{,1}&=&
\rho\left[(\psi_{,1})^2-(\psi_{,2})^2\right]
-\frac{1}{4\rho}\left[(\omega_{,1})^2-(\omega_{,2})^2\right]
\e^{4\psi} \nonumber \\
& & -\rho(\Phi_{,1}\bar{\Phi}_{,1}-\Phi_{,2}\bar{\Phi}_{,2})\e^{-2\psi} 
    +\rho\left[(\varphi_{,1})^2-(\varphi_{,2})^2\right],
\label{a10} \\
\gamma_{,2}&=&
2\rho\psi_{,1}\psi_{,2}
-\frac{1}{2\rho}\omega_{,1}\omega_{,2}\e^{4\psi}
-\rho(\Phi_{,1}\bar{\Phi}_{,2}+\Phi_{,2}\bar{\Phi}_{,1})\e^{-2\psi}
+2\rho\varphi_{,1}\varphi_{,2},
\label{a11} \\
\nabla^2\varphi & = & 0, 
\label{a12}
\end{eqnarray}
\end{subequations}
where, for any functions, $f$ and $h$,
\begin{equation}
\nabla^2 f\equiv\frac{1}{\rho}\frac{\partial}{\partial\rho}
\left(\rho\frac{\partial f}{\partial\rho}\right)
+ \frac{\partial^2 f}{\partial z^2},~~~
\nabla f\cdot\nabla h\equiv
\frac{\partial f}{\partial \rho}\frac{\partial h}{\partial \rho}+
\frac{\partial f}{\partial z}\frac{\partial h}{\partial z},~~~
f_{,i}\equiv\frac{\partial f}{\partial x^i}.
\end{equation}
In Ref.\cite{21},
one will find the explicit forms of the metric function, $\omega$, and
the Maxwell field in terms of the complex potentials, 
${\cal E}$ and $\Phi$. 
Note that the first, second and third equations are the same as those in
general relativity and are referred to as 
{\it the Ernst equations}
of the Einstein-Maxwell system. 
The remaining equations contain the scalar field contributions,
and, as for the metric functions, 
effects of the scalar field only appear
in $\gamma$.

\

\section{Scalar-tensor-Weyl solutions.}

\subsection{The reduced Ernst equation}

\

In this section, we shall consider static, axisymmetric
vacuum solutions.
In this case, $\omega=\Phi=0$, and
the field equations (\ref{a7})$\sim$(\ref{a12})
are reduced to the following:
\begin{subequations}
\begin{eqnarray}
\e^{2\psi} &=& \EP,\\
\EP\nabla^2{\EP} & = & (\nabla\EP)\cdot(\nabla\EP), 
\label{b2} \\
\gamma_{,1}&=&
\rho\left[(\psi_{,1})^2-(\psi_{,2})^2\right]
+\rho\left[(\varphi_{,1})^2-(\varphi_{,2})^2\right],
\label{b3} \\
\gamma_{,2}&=&
2\rho\psi_{,1}\psi_{,2}
+2\rho\varphi_{,1}\varphi_{,2},
\label{b4} \\
\nabla^2\varphi & = & 0,
\label{b5}
\end{eqnarray}
\end{subequations}
where ${\EP}$ is a real function.
One finds that the Ernst equation (\ref{b2}) is reduced to
\begin{equation}
\nabla^2\psi = 0.
\label{b7}
\end{equation}
That is, $\varphi$ and $\psi$ are harmonic functions.

We introduce {\it oblate} and {\it prolate} coordinates, $(x,y)$, 
defined by
\begin{subequations}
\begin{eqnarray}
\rho & =& \sigma \sqrt{(x^2+\epsilon)(1-y^2)} , 
\label{b8} \\
z & = & \sigma xy ,
\label{b9}
\end{eqnarray}
\end{subequations}
where $\sigma$ is a positive constant, and $\epsilon=\pm 1$.
The cases, $\epsilon=1$ and $\epsilon=-1$, are referred to as
{\it oblate} and {\it prolate}, respectively. 
The Ernst equation is then reduced to
\begin{equation}
\frac{\partial}{\partial
  x}\left[(x^2+\epsilon)\frac{\partial\psi}{\partial x}\right]
+\frac{\partial}{\partial
  y}\left[(1-y^2)\frac{\partial\psi}{\partial y}\right] =0 .
\label{b10}
\end{equation}
The similar equation holds for $\varphi$.

\

\subsection{Prolate solutions}

\

The simplest prolate solution for $\psi$ is given by
\begin{equation}
\psi=\frac{\delta}{2}\ln\left(\frac{x-1}{x+1}\right),
\label{b11}
\end{equation}
where $\delta$ is an integration constant.
The similar solution for $\varphi$ is given by
\begin{equation}
\varphi = \varphi_0 + \frac{d}{2}\ln\left(\frac{x-1}{x+1}\right) ,
\label{b13}
\end{equation}
where $\varphi_0$ and $d$ are integration constants.
Then the corresponding solution for $\gamma$ becomes
\begin{equation}
\e^{2\gamma}=\left(\frac{x^2-1}{x^2-y^2}\right)^{{\Delta}^2} ,
\label{b14}
\end{equation}
where 
\begin{equation}
{\Delta}^2\equiv \delta^2 +d^2 .
\label{b15}
\end{equation}
For completeness, an explicit form of the metric is
shown below:
\begin{eqnarray}
ds^2 & = & -\left(\frac{x-1}{x+1}\right)^{\delta} dt^2 
+\sigma^2 \left(\frac{x-1}{x+1}\right)^{-\delta} \times \nonumber \\
& & 
\left[\left(\frac{x^2-1}{x^2-y^2}\right)^{\Delta^2}
(x^2-y^2)\left(\frac{dx^2}{x^2-1}+\frac{dy^2}{1-y^2}\right)+(x^2-1)(1-y^2)
d\phi^2\right].
\label{b16}
\end{eqnarray}

Note that, though $\psi$ and $\varphi$ are $y$-independent,
$\gamma$ depends on both $x$ and $y$.
The metric contains two parameters, $\delta$ and $\Delta$.
Now we show that the metric is reduced to
the previously known one when we take specific values of the parameters.
When $\Delta=\delta$, i.e., $d=0$, 
the metric is reduced to one of the Weyl series of the solutions
in general relativity (Voorhees's prolate solution \cite{22}).
Therefore, we refer to (\ref{b16}) as the scalar-tensor-Weyl
solution. 
Voorhees's prolate solution contains the Schwarzschild solution
as a specific case, $\delta=1$, and the  
Schwarzschild coordinates are related to $(x,y)$ as
\begin{equation}
x = \frac{r}{m}-1, \ \ y=\cos\theta, \ \ \sigma=m,
\label{b30}
\end{equation}
where $m$ is a usual mass parameter.

The case, $\Delta=1$, is the most interesting one.
We introduce new parameters, $a$, $b$ and $c$,
and new coordinates (the Just coordinates) by
\begin{equation}
d = \frac{2c}{a}, \ \ \sigma=\frac{a}{2}, \ \ 
\delta = \frac{b}{a} , \label{b20} 
\end{equation}
and 
\begin{equation}
x = \frac{2\chi}{a}-1, \ \
y = \cos\theta \label{b22}.
\end{equation}
Then we find that the solution coincides with
the spherical exterior solution with a vanishing electric charge:
\begin{equation}
ds^2 = -\left(1-\frac{a}{\chi}\right)^{\frac{b}{a}}dt^2
+\left(1-\frac{a}{\chi}\right)^{-\frac{b}{a}}d\chi^2
+\chi^2\left(1-\frac{a}{\chi}\right)^{\frac{a-b}{a}}
\left(d\theta^2+\sin^2\theta d\phi^2\right).
\label{ses}
\end{equation}
Moreover, when $\delta=1$, i.e., $a=b$, 
the solution is reduced to
Schwarzschild's one.

The right hand sides in (\ref{b3}) and (\ref{b4})
are in the forms of the superposition of
the contributions of $\psi$ and $\varphi$.
It is important to note that 
the superposition of 
the {\it effectively non-spherical}
$\psi$-terms corresponding to Voorhees's prolate solution with $\delta\neq1$
and the {\it spherical} configuration of $\varphi$ 
reduces to the solution containing the spherical exterior solution.
When the spherical scalar field configuration is added to
the $\psi$-terms corresponding 
to Voorhees's prolate solution with $\delta=1$,
namely the Schwarzschild solution, we cannot have 
spherical solutions unless $a=b$. The relation among these
solutions is schematically shown in Figure 1.
The scalar-tensor-Weyl solution generically has 
a singularity, $x=1$, whose topology is $S^2$,
and qualitative features of the singularity may be similar to
those in Voorhees's prolate solution.
When $\Delta=1$, the singularity, $x=1$, becomes a point,
which is given $\chi=a$ in the Just coordinate.
Moreover, when $\Delta=\delta=1$, the topology of
$\chi=a$, corresponding to the event horizon, $r=2m$, becomes $S^2$.

\begin{figure}
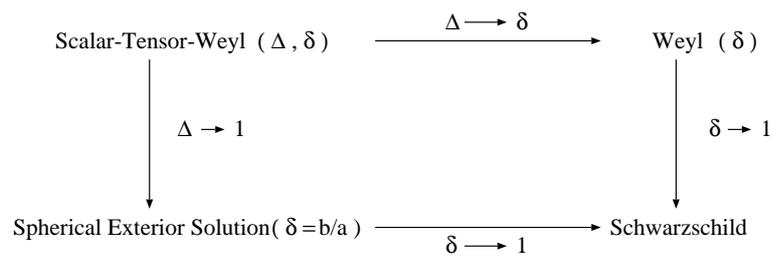

\begin{center}
\hspace*{0.5cm}
\caption{A schematic picture of series of 
the scalar-tensor-Weyl solutions.}
\label{fig}
\end{center}
\end{figure}

\

\subsection{Oblate solutions}

\

The simplest oblate solution for $\psi$ is given by
\begin{equation}
\psi=\delta\cdot\mbox{arccot}{x} ,
\label{b40}
\end{equation}
where $\delta$ is an integration constant.
The similar solution for $\varphi$ is given by
\begin{equation}
\varphi = \varphi_0 + d\cdot\mbox{arccot}x, 
\label{b42}
\end{equation}
where $\varphi_0$ and $d$ are integration constants.
Then the corresponding solution for $\gamma$ becomes
\begin{equation}
\e^{2\gamma}=\left(\frac{x^2+y^2}{x^2+1}\right)^{{\Delta}^2} ,
\label{b43}
\end{equation}
where ${{\Delta }^2}=\delta^2+d^2$.
For completeness, an explicit form of the metric is
shown below:
\begin{eqnarray}
ds^2 & = & - \e^{2\delta\cdot\mbox{\scriptsize arccot}x} dt^2 
+\sigma^2 \e^{-2\delta\cdot\mbox{\scriptsize arccot}x} \times \nonumber \\
& &
\left[\left(\frac{x^2+y^2}{x^2+1}\right)^{\Delta^2}
(x^2+y^2)\left(\frac{dx^2}{x^2+1}+\frac{dy^2}{1-y^2}\right)+(x^2+1)(1-y^2)
d\phi^2\right].
\label{b45}
\end{eqnarray}
When $\Delta=\delta$, 
the metric is reduced to Voorhees's oblate solution.
In contrast to the prolate case, the oblate
solution is not  reduced to 
the spherical exterior solutions for any $\delta$ and $\Delta$.

\

\newpage
\section{Local behaviors of null geodesics}

\subsection{The spherical exterior solution}

\

In this section, we examine local behaviors 
of irrotational null geodesics
in the spherical exterior solution
with a vanishing electric charge.
The metric is rewritten in the Schwarzschild coordinate as
\begin{equation}
\eqalign{
&ds^2=-\left(1-\frac{a}{\chi(r)}\right)^{\frac{b}{a}}dt^2
+\left(1-\frac{2m(r)}{r}\right)^{-1}dr^2+r^2d\Omega^2,\cr
&2m(r)=\left(b-\frac{(a+b)^2}{4\chi(r)}\right)
\left(1-\frac{a}{\chi(r)}\right)^{-\frac{(a+b)}{2a}},~~~~~
r=\chi(r)\left(1-\frac{a}{\chi(r)}\right)^{\frac{(a-b)}{2a}},
\cr
}
\label{ge1}
\end{equation}
where the mass function, $m(r)$, can be interpreted as
a local energy \cite{89}.
One immediately finds that $m(r)$ becomes negative
when $(a+b)^2/4b>\chi>a$. 
Therefore, one may expect that there will be
significant difference not only in the global causal structures
but also in the local geometry, 
depending on whether $a$ is equal to $b$ or not.
However, we will find that it is not the case and that
local behaviors of null geodesics significantly depend on
whether $a$ is greater than $2b$ or not.

Due to the conformally invariant nature of null geodesics,
we consider null geodesics in the spacetime
with a metric, ${\bf g}\equiv A^{-2}\hat{\bf g}$ (see Appendix A).
Let $\hat{k}^\mu=dx^\mu/dv$ be
null geodesic associated with $\hat{\bf g}$
with the affine parameter, $v$. 
Hereafter, a hat denotes
geometrical quantities associated with $\hat{\bf g}$,
and the corresponding geometrical quatities associated with ${\bf g}$
are denoted without a hat.
Then the null geodesics associated with ${\bf g}$ are given as
\begin{equation}
d\lambda=A^{-2}dv,~~k^\mu=\frac{dx^\mu}{d\lambda}=A^2\hat{k}^\mu,
~~~~\rightarrow~~~~~
\hat{k}^\alpha\hat{\nabla}_\alpha\hat{k}^\mu
=
k^\alpha\nabla_\alpha k^\mu=0.
\label{ge2}
\end{equation}
Since the local geometrical nature of the spacetime is
described by the Riemann curvature, it is important to
examine optical scalars of the irrotational null geodesic congruence
defined by Sachs \cite{90}.
Let $\{\hat{\bf E}_{(a)}:~(a=1\sim4)\}
=\{\hat{\bf k}, \hat{\bf m}, \hat{\bf t}, \bar{\hat{\bf t}}\}$
be a null tetrad such that
\begin{equation}
\hat{k}^\alpha\hat{\nabla}_\alpha\hat{E}_{(a)}^\mu=0,~~~
\hat{g}_{\mu\nu}\hat{E}_{(a)}^\mu\hat{E}_{(b)}^\nu=
-2\delta^1_{(a}\delta^2_{b)}-
2\delta^3_{(a}\delta^4_{b)}.
\label{ge3}
\end{equation}
Then the optical scalars, namely, the expansion, $\theta$, and
the complex shear, $\sigma$, are defined as  
\begin{equation}
\hat{\theta}=\hat{\nabla}_\mu\hat{k}_\nu\hat{t}^\mu
\bar{\hat{t}^\nu},~~~
\hat{\sigma}=\hat{\nabla}_\mu\hat{k}_\nu\bar{\hat{t}^\mu}
\bar{\hat{t}^\nu}.
\label{ge4}
\end{equation}
The optical scalar equations are also conformally invariant,
and we define the following quantities associated with ${\bf g}$:
\begin{equation}
\theta=A^2\hat{\theta}-k^\mu\nabla_\mu\ln A,~~
\sigma=A^2\hat{\sigma},~~t^\mu=A\hat{t}^\mu.
\label{ge5}
\end{equation}
Then the optical scalar equations become \cite{90,91}
\begin{equation}
\eqalign{
&\frac{d\theta}{d\lambda}+\theta^2+|\sigma|^2
=-\frac{1}{2}R_{\mu\nu}k^\mu k^\nu\equiv{\cal R},\cr
&\frac{d\sigma}{d\lambda}+2\theta\sigma
=-R_{\mu\alpha\nu\beta}k^\mu k^\nu\bar{t}^\alpha \bar{t}^\beta
=-C_{\mu\alpha\nu\beta}k^\mu k^\nu\bar{t}^\alpha \bar{t}^\beta
\equiv F,
\cr}
\label{ge6}
\end{equation}
where $C_{\mu\alpha\nu\beta}$ is the Weyl curvature.

First, we examine circular orbits
of null geodesics in the Just coordinate.
Without loss of generality, we consider null geodesics
on the $\theta=\pi/2$ plane. Then 
the geodesic equation is reduced to
\begin{equation}
\eqalign{
&
\dot{\chi}^2=E^2
-\frac{L^2}{\chi^2}\left(1-\frac{a}{\chi}\right)^{\frac{(2b-a)}{a}}
\equiv E^2-V(\chi),\cr
&E=\left(\aoverchi\right)^{\frac{b}{a}}\dot{t},~~~
L=\chi^2\left(\aoverchi\right)^{\frac{(a-b)}{a}}\dot{\phi},\cr}
\label{ge7}
\end{equation}
where a dot denotes a derivative with respect to
the affine parameter, $\lambda$, and $E$ and $L$ are integration constants.
The circular orbits are determined by the conditions that
$\dot{\chi}=0$ and $dV/d\chi=0$. We find that the circular orbit is
obtained as $\chi=\chi_C\equiv(a+2b)/2$ for $a<2b$, and that, 
surprisingly, no circular orbit 
exists for $a\geq2b$ in contrast to the case in the Schwarzschild
spacetime.

Dyer \cite{91} has obtained general solutions to the optical scalar
equations $(\ref{ge6})$ in the general static, spherically
symmetric spacetime, and we summarize his results in Appendix B. 
In the static, spherically symmetric spacetime,
one finds that the Weyl driving term, $F$, 
and the shear, $\sigma$, can be regarded as real quantities
without loss of generality \cite{91}.
By defining the new optical scalars, $C_\pm$, as
\begin{equation}
\frac{d}{d\lambda}\ln C_\pm=\theta\pm\sigma,
\label{ge8}
\end{equation}
Dyer has obtained the following equations:
\begin{equation}
\frac{d^2C_\pm}{d\lambda^2}=({\cal R}\pm F)C_\pm.
\label{ge9}
\end{equation}

We evaluate the Ricci and Weyl driving terms, ${\cal R}$ and $F$,
as
\begin{equation}
\eqalign{
{\cal R}&=-\frac{(a^2-b^2)}{4\chi^4}\left(\aoverchi\right)^{-2}
\left\{1-\frac{h^2}{\chi^2}
\left(\aoverchi\right)^{\frac{(2b-a)}{a}}\right\}
\cr
&=-\frac{(a^2-b^2)}{4\chi^4}\left(\aoverchi\right)^{-2}\left(k^1\right)^2,\cr
F&=\frac{h^2}{4\chi^6}\left(\aoverchi\right)^{\frac{(2b-3a)}{a}}
\left\{6b\chi-(a+b)(a+2b)\right\},
\cr
}
\label{ge10}
\end{equation}
where $h=L/E$ is an impact parameter, 
and the affine parameter, $\lambda$, is chosen such that $E=1$.
It is immediately found that the Ricci driving term, ${\cal R}$,
vanishes at the perihelion where $k^1=0$.
A particularly interesting case is the circular orbit
on which $k^1=0$ identically. 
If $F$ vanishes on the circular
orbit, the image shape of an infinitesimal light ray congruence
remains unchanged. The condition, $F=0$, is reduced to
$\chi=\chi_F\equiv(a+b)(a+2b)/(6b)$ for $a>2b$. 
When $a<2b$, $F$ is strictly positive.
Since the circular orbit exists only when $a<2b$,
the condition, ${\cal R}=F=0$, is never satisfied on the
circular orbit.
However, the condition, ${\cal R}=F=0$, can be satisfied 
at $\chi=\chi_F$ for the scattering
orbit when the impact parameter, $h$, is chosen as
\begin{equation}
h=\frac{(a+b)(a+2b)}{6b}\left[\frac{(a-b)(a-2b)}{(a+b)(a+2b)}
\right]^{\frac{(a-2b)}{2a}},
~~~~\rightarrow~~~~~{\cal R}(\chi_F)=F(\chi_F)=0.
\label{ge11}
\end{equation}
That is, {\it the gravity effectively vanishes at $\chi_F$} in this sense.

In summary, we have found that
local behaviors of null geodesics significantly depend on
whether $a$ is greater than $2b$ or not.
When $a<2b$, the behaviors are similar to those in 
the Schwarzschild spacetime. That is, the circular orbit exists,
and the Weyl driving term, $F$, is strictly positive.
When $a>2b$, on the contrary, there is no circular orbit,
and $F$ changes its sign at $\chi_F$.

These behaviors of null geodesics can be compared with those
in the Reissner-Nordstr\"om spacetime:
\begin{equation}
ds^2=-\left(1-\frac{2m}{r}+\frac{Q^2}{r^2}\right)dt^2
+\left(1-\frac{2m}{r}+\frac{Q^2}{r^2}\right)^{-1}dr^2
+r^2d\Omega^2.
\label{ge12}
\end{equation}
The Ricci and Weyl driving terms are calculated as
\begin{equation}
{\cal R}=-\frac{Q^2h^2}{r^4},~~~
F=\frac{3h^2}{r^4}\left(\frac{m}{r}-\frac{Q^2}{r^2}\right).
\label{ge13}
\end{equation}
One immediately finds that $F$ vanished at
$r=r_F\equiv Q^2/m$. However,
when $m^2>Q^2$, $r_F$ is
inside the event horizon, $r_+\equiv m+\sqrt{m^2-Q^2}$.
When $9m^2>8Q^2$, two circular orbits exist:
\begin{equation}
r=r_C^\pm\equiv\frac{1}{2}\left(3m\pm\sqrt{9m^2-8Q^2}\right).
\label{ge14}
\end{equation}
Moreover, when $m^2>Q^2$, one finds that $r_C^+>r_+>r_F>r_C^->r_-$, 
where $r_-\equiv m-\sqrt{m^2-Q^2}$. 
A particular case that $r_F=r_C^-$ occurs only when $m^2=Q^2$,
and one finds that $r_F=r_C^-=m=r_\pm$ and that $r_C^+=2m$. 

General solutions to $(\ref{ge9})$ with $(\ref{ge10})$
can be obtained by Dyer's formula (Appendix B.1),
however, we have only examined quantitative behaviors
of the Ricci and Weyl driving terms, ${\cal R}$ and $F$.
Further quantitative studies of the solutions and
their astronomical applications, especially
the implication to gravitational lens effects, 
will be discussed in the forthcoming paper.

\

\subsection{The scalar-tensor-Weyl solution}

\

It is difficult to study generic
null geodesics in the scalar-tensor-Weyl solution.
Therefore, we shall examine 
specific null geodesics on the $y=0$ plane.
When $y=0$, the geodesic equation is reduced to 
\begin{equation}
\begin{array}{l}
\dot{W}^2=E^2-\dfrac{L^2}{x^2-1}\left(\dfrac{x-1}{x+1}\right)^{2\delta}
\equiv E^2-V(x), \\
E=\dfrac{1}{\sigma}\left(\dfrac{x-1}{x+1}\right)^{\delta}\dot{t}, \ \ \ 
L=(x^2-1)\left(\dfrac{x-1}{x+1}\right)^{-\delta}\dot{\phi},
\end{array}
\label{n1}
\end{equation}
where $W$ is a function of $x$ defined by
\begin{equation}
\dfrac{dx}{dW}=\left(\dfrac{x^2}{x^2-1}\right)^{\frac{\Delta^2-1}{2}},
\label{nx}
\end{equation}
a dot denotes a derivative with respect to the affine parameter, 
$\lambda$, and $E$ and $L$ are integration constants. 
The circular
orbits are determined by the conditions that $\dot{W}=0$ and
$dV/dx=0$. We find that these conditions are reduced to
$x=x_{\mbox{\tiny C}}\equiv2\delta$.
Since $x$ should be larger than unity, 
the circular orbit exists when $\delta>1/2$.

We evaluate the Ricci and Weyl deriving terms, ${\cal R}$ and $F$,
(see Appendix B.2) and find that
\begin{equation}
\begin{array}{rcl}
{\cal R} & = &
-\dfrac{\Delta^2-\delta^2}{\sigma^2}
\dfrac{1}{x^2(x^2-1)}\left(\dfrac{x^2}{x^2-1}
\right)^{\Delta^2}\left[1-\dfrac{(h/\sigma)^2}{x^2-1}
\left(\dfrac{x-1}{x+1}\right)^{2\delta}\right] \\
 & = & -\dfrac{\Delta^2-\delta^2}{(x^2-1)^2}(k^1)^2 , \\
F & = &\dfrac{1}{\sigma^2}\dfrac{1}{x^3(x^2-1)}\left(\dfrac{x^2}{x^2-1}
\right)^{\Delta^2} \times \\
& & \times \left[\delta(\Delta^2-1)+\dfrac{(h/\sigma)^2}
{x^2-1} \left(\dfrac{x-1}{x+1}\right)^{2\delta}
\left\{\delta(\Delta^2-1)-(\Delta^2+2\delta^2)x+3\delta x^2\right\}
\right],
\end{array}
\label{n2}
\end{equation}
where $h=L/E$ is an impact parameter.
One finds that 
they are reduced to (\ref{ge10}) when $\Delta^2=1$.
We show $\cal{R}$ and $F$ in Figure 2.
We numerically find that the qualitative features of ${\cal R}$
and $F$ are the same as those in the spherical case ($\Delta=1$)
and that the significant qualitative changes in 
$\cal{R}$ and $F$ are seen depending on whether $\delta<1/2$ or not.


\
\section{Discussions}

\

After summarizing the spherical exterior solutions
in the scalar-tensor theories of gravity,
we have derived a new series of the axisymmetric exterior
solutions, which are reduced
to Voorhees's solutions in general relativity.
We have found that the prolate solution,
namely the scalar-tensor-Weyl solution,
contains the spherical exterior solution as a special case in which
we take some non-trivial choice of the parameters
in the solution.
We also tried to find exterior solutions corresponding
to the Kerr solution in general relativity
and have not yet succeeded in it.
Note that the generic scalar-tensor-Weyl solution
is not reduced to the Schwarzschild solution
in the limit, $\varphi\rightarrow0$.
Therefore, we guess that the Kerr-like exterior solution
must be reduced to the Tomimatsu-Sato-like solution
in the limit, $\varphi\rightarrow0$,
where, by the term, the Kerr-like exterior solution,
we mean the exterior solution such that
it coincides with the Kerr solution
when $\varphi=0$ and that it is reduced to
the spherical exterior solution when the Kerr parameter
(the angular momentum of the matter) vanishes.
An example of the solution with the first property is obtained
by using the metric function, $\psi$, corresponding to
the Kerr solution. However, the resulting solution
does not satisfy the second property.
We think that generating methods of the exact solutions
(e.g. the Newman-Janis algorithm \cite{98})
may play an important role in obtaining
further exterior solutions.

It is important to note that
one have to specify an explicit form of
the coupling function, $A(\varphi)$, in order to
obtain the complete
information, especially the global geometry,
of the exterior solutions in the scalar-tensor theories.
We have not given any discussion on $A(\varphi)$,
and, instead, we have examined local geometrical properties
of the solutions by taking account of
the conformally invariant nature of null geodesics.
We have evaluated the Ricci and Weyl deriving terms,
${\cal R}$ and $F$, in the scalar-tensor-Weyl solution
and have found that
local behaviors of null geodesics significantly depend on
whether the parameter, $\delta$, which corresponds to
$b/a$ in the spherical exterior solution with $Q=0$,
is greater than $1/2$ or not.
When $\delta>1/2$, the behaviors are similar to those in
the Schwarzschild spacetime. That is, the circular orbit exists,
and the Weyl driving term, $F$, is strictly positive.
When $\delta<1/2$, on the contrary, there is no circular orbit,
and $F$ changes its sign at $\chi_F$.
In the later case, one may expect to observe
a gravitational lensing anomaly for images of distant sources
appearing near $\chi_F$. A ratio, 
\begin{equation}
{\cal C}(h) \equiv
\lim_{\lambda\rightarrow\infty}\dfrac{C_+-C_-}{C_+} ,
\end{equation}
is a measure of the image deformation.
In the Schwarzschild solution,
${\cal C}$ is a positive, decreasing function of $h$.
However, in the spherical exterior solution with $\delta<1/2$,
${\cal C}$ may become a negative increasing function
of $h$ near $\chi_F$.
Further quantitative studies of the solutions, $C_+$ and $C_-$,
and their astronomical applications, especially
the implication to gravitational lens effects
including their observational detectability,
will be discussed in the forthcoming paper.

\

\appendix
\section{Summary of the scalar-tensor theories}
\label{ap1}

\

We shall consider the simplest scalar-tensor theory
[2-5,14].  In this theory, gravitational
interactions are mediated by a tensor field, $ \hat{g}_{\mu\nu} $, and
a scalar field, $ \hat{\phi} $.  Hereafter, a symbol, $\hat{}$, denotes
quantities or derivatives associated with $\hat{g}_{\mu\nu}$.  An
action of the theory is the following:
\begin{equation}
  S=\frac{1}{16\pi}\int
    \left[\hat{\phi}
    \hat{R}-\frac{\omega(\hat{\phi})}
    {\hat{\phi}}\hat{g}^{\mu\nu}\hat{\phi}_{,\mu}\hat{\phi}_{,\nu}
    \right]\sqrt{-\hat{g}}d^4x +
  S_{\mbox{\tiny matter}}[\hat{\Psi}_{\mbox{\tiny
      m}},\hat{g}_{\mu\nu}],
\label{ea5}
\end{equation}
where $\omega(\hat{\phi})$ is a dimensionless arbitrary 
function of $\hat{\phi}$,
$\hat{\Psi}_{\mbox{\tiny m}}$ represents matter fields, and
$S_{\mbox{\tiny matter}}$ is an action of the matter fields.  The
scalar field, $\hat{\phi}$, plays a role of an effective gravitational
constant as $\hat{G}\sim1/\hat{\phi}$.  Varying the action by the tensor
field, $\hat{g}_{\mu\nu}$, and the scalar field, $\hat{\phi}$, yields,
respectively, the following field equations:
\begin{eqnarray}
  \hat{G}_{\mu\nu}& = & \frac{8\pi}{\hat{\phi}}\hat{T}_{\mu\nu}
  +\frac{\omega(\hat{\phi})}{\hat{\phi}^2}
  \left(\hat{\phi}_{,\mu}\hat{\phi}_{,\nu}-\frac{1}{2}\hat{g}_{\mu\nu}
    \hat{g}^{\alpha\beta}\hat{\phi}_{,\alpha}
    \hat{\phi}_{,\beta}\right)+
    \frac{1}{\hat{\phi}}(\hat{\nabla}_{\mu}\hat{\nabla}_{\nu}\hat{\phi}
  -\hat{g}_{\mu\nu} {\hat{\scriptstyle\Box}}\hat{\phi}), \label{ea6} \\ 
{\hat{\scriptstyle\Box}}\hat{\phi} & =
  & \frac{1}{3+2\omega(\hat{\phi})}\left(8\pi \hat{T} -
    \frac{d\omega}{d\hat{\phi}}
    \hat{g}^{\alpha\beta}\hat{\phi}_{,\alpha}\hat{\phi}_{,\beta}\right).
\label{ea7}
\end{eqnarray}

Now we perform the following conformal transformation
to the unphysical frame (the Einstein frame) with the metric,
${\bf g}$:
\begin{equation}
  g_{\mu\nu}=A^{-2}(\varphi)\hat{g}_{\mu\nu},
\label{ea8}
\end{equation}
such that
\begin{equation}
  GA^2(\varphi) = \frac{1}{\hat{\phi}},
\label{ea9}
\end{equation}
where $G$ is a bare gravitational constant, 
and $A(\varphi)$ is referred to as a coupling function.
Then the action is rewritten as
\begin{equation}
  S=\frac{1}{16\pi
    G}\int
  (R-2g^{\mu\nu}\varphi_{,\mu}\varphi_{,\nu})\sqrt{-g}d^4x +
  S_{\mbox{\tiny matter}}[\hat{\Psi}_{\mbox{\tiny m}},A^2(\varphi)
  g_{\mu\nu}],
\label{ea10}
\end{equation}
where the scalar field, $\varphi$, is defined by
\begin{equation}
  \alpha^2(\varphi) \equiv \left(\frac{d\ln
      A(\varphi)}{d\varphi}\right)^2 = \frac{1}{3+2\omega(\hat{\phi})}.
\label{ea11}
\end{equation}
Varying the action by $g_{\mu\nu}$ and $\varphi$ yields,
respectively,
\begin{eqnarray}
  G_{\mu\nu}& = & 8\pi G T_{\mu\nu} +
  2\left(\varphi_{,\mu}\varphi_{,\nu}
    -\frac{1}{2}g_{\mu\nu}g^{\alpha\beta}\varphi_{,\alpha}
    \varphi_{,\beta}\right),
\label{ea12} \\
\Box\varphi & = &-4\pi G\alpha(\varphi) T ,
\label{fs2}
\end{eqnarray}
where $T^{\mu\nu}$ represents a energy-momentum tensor with respect to
$g_{\mu\nu}$ defined by
\begin{equation}
  T^{\mu\nu} \equiv \frac{2}{\sqrt{-g}} \frac{\delta S_{\mbox{\tiny
        matter}}[\hat{\Psi}_{\mbox{\tiny m}},A^2(\varphi)g_{\mu\nu}]}
  {\delta g_{\mu\nu}}=A^6(\varphi)\hat{T}^{\mu\nu}.
\label{ea14}
\end{equation}
The conservation law of $T^{\mu\nu}$ is given by
\begin{equation}
  \nabla_{\nu}T^{\, \mu}_{\nu}=\alpha(\varphi)T \nabla_{\mu}\varphi.
\label{ea15}
\end{equation}
For the Maxwell field, one has
\begin{equation}
   T^{\mu\nu}=\frac{1}{4\pi}\left(
F_{\mu\alpha}F_\nu^{~\alpha}
-\frac{1}{4}g_{\mu\nu}F_{\alpha\beta}F^{\alpha\beta}
\right).
\label{max}
\end{equation}
Further properties of the Maxwell field under the conformal
transformation are found in Ref.\cite{20}.

In this paper, we adopt the unit, $G=1$.

\

\section{Analytic solutions to the optical scalar equations}
\label{ap10}

\subsection{Static, spherically symmetric spacetime}

\

In this Appendix,
we summarize the analytic results obtained by Dyer \cite{91}.
A metric of
the static, spherically symmetric spacetime is given as
\begin{equation}
ds^2=-\e^{2C}dt^2+\e^{2A}dr^2+\e^{2B}d\Omega^2,
\label{gap1}
\end{equation}
where $A$, $B$ and $C$ are functions of $r$.
The null tangent vector, $k^\mu$, to a null geodesic
in this spacetime is obtained as
\begin{equation}
\eqalign{
&k^0=\e^{-2C},~~
k^2=0,~~k^3=h\e^{-2B},\cr
&k^1=\pm\e^{-(A+C)}\sqrt{1-h^2\e^{2(C-B)}},\cr}
\label{gap2}
\end{equation}
where a constant, $h$, is an impact parameter, and
we assume that the geodesic is on the $\theta=\pi/2$ plane
without loss of generality. 
The complex null vector, $t^\mu$, is obtained as
\begin{equation}
\eqalign{
&t^0=\frac{\e^{-C}}{2\sqrt{2}}\left[\frac{2}{h}S_+S_-\e^{B}
+H(S_+^2+S_-^2)\right],~~t^2=\frac{i}{\sqrt{2}}\e^{-B},\cr
&t^1=\frac{\e^{-A}}{2\sqrt{2}}\left[\frac{1}{h}(S_+^2+S_-^2)\e^{B}
+2HS_+S_-\right],~~t^3=\frac{Hh}{\sqrt{2}}\e^{-2B},\cr
&S_\pm=\sqrt{\e^{-C}\pm h\e^{-B}},~~
H=-\frac{1}{h}\int^r\frac{B'}{S_+S_-}\e^{B-C}dr,
}
\label{gap3}
\end{equation}
where a prime denotes a derivative with respect to $r$.
Then the Ricci and Weyl driving terms, ${\cal R}$ and $F$, 
in $(\ref{ge6})$ are evaluated as
\begin{equation}
\eqalign{
&{\cal R}+F=\e^{-2(A+C)}\left(B''+(B')^2-B'C'-B'A'\right)\cr
&~~~~~~~~~~~-h^2\e^{-2(A+B)}\left(C''+(C')^2-C'B'-C'A'\right),
\cr
&{\cal R}-F=\e^{-2(A+C)}\left(B''+(B')^2-B'C'-B'A'\right)\cr
&~~~~~~~~~~~-h^2\e^{-2(A+B)}\left(\e^{2(A-B)}+B''-A'B'\right).
\cr}
\label{gap4}
\end{equation}
Since $F$ is real, one can let $\sigma$ be real without
loss of generality.
Dyer has introduced the new optical scalars, $C_\pm$, defined by
\begin{equation}
\frac{d}{d\lambda}\ln C_\pm=\theta\pm\sigma,
\label{gap5}
\end{equation}
and has obtained the following equations:
\begin{equation}
\frac{d^2C_\pm}{d\lambda^2}=({\cal R}\pm F)C_\pm.
\label{gap6}
\end{equation}
Dyer has obtained general solutions to $(\ref{gap6})$ as
\begin{equation}
\eqalign{
&C_+=C_+^0\sqrt{\e^{2B}-h^2\e^{2C}}\left\{\int^{r}
\frac{\e^{A+B+C}}{\left[\e^{2B}-h^2\e^{2C}\right]^{\frac{3}{2}}}dr
+D_+^0\right\},
\cr
&C_-=C_-^0\e^{B}\sin\left(h
\int^r \frac{\e^{A+C}}{\sqrt{\e^{2B}-h^2\e^{2C}}}dr
+D_-^0\right),
\cr
}
\label{gap7}
\end{equation}
where $C_\pm^0$ and $D_\pm^0$ are integration constants.

\

\subsection{Static, axisymmetric spacetime}\label{ap5}

\

A metric of the static, axisymmetric spacetime
is given by
\begin{equation}
ds^2=-\e^{2\alpha}dt^2+\e^{2\beta}dx^2+\e^{2\gamma}dy^2+\e^{2\mu}d\phi^2,
\label{k1}
\end{equation}
where metric functions, $\alpha$, $\beta$, $\gamma$ and $\mu$,
are functions of $x$ and $y$.
The tangent vector, $k^\mu$, to a null geodesic on the $y=0$ plate is
obtained as 
\begin{equation}
\begin{array}{l}
k^0=\e^{-2\alpha}, \ \ \ k^2=0, \ \ \ k^3=h\e^{-2\mu}, \\
k^1=\pm \e^{-\beta}\sqrt{\e^{-2\alpha}-h^2\e^{-2\mu}},
\end{array}
\end{equation}
where a constant, $h$, is an impact parameter.
The complex null vector, $t^\mu$, is obtained as 
\begin{equation}
\eqalign{
&t^0=\frac{1}{2\sqrt{2}}\left[\frac{2}{h}\e^{\mu}S_{+}S_{-}
+H(S_{+}^2+S_{-}^2)\right]\e^{-\alpha},~~ 
t^2=\frac{i}{\sqrt{2}}\e^{-\gamma}, \cr
&t^1=\frac{1}{2\sqrt{2}}\left[\frac{1}{h}(S_{+}^2+S_{-}^2)\e^{\mu}
+2HS_{+}S_{-}\right]\e^{-\beta},~~
 t^3=\frac{1}{\sqrt{2}}Hh\e^{-2\mu},\cr
&S_{\pm} = \sqrt{\e^{-\alpha}\pm h\e^{-\mu}}, ~~
H=-\frac{1}{h}\int^x
\frac{\e^{\mu-\alpha}}{S_{+}S_{-}}
\frac{\partial\gamma}{\partial x}dx.\cr
}
\end{equation}

\



\begin{figure}
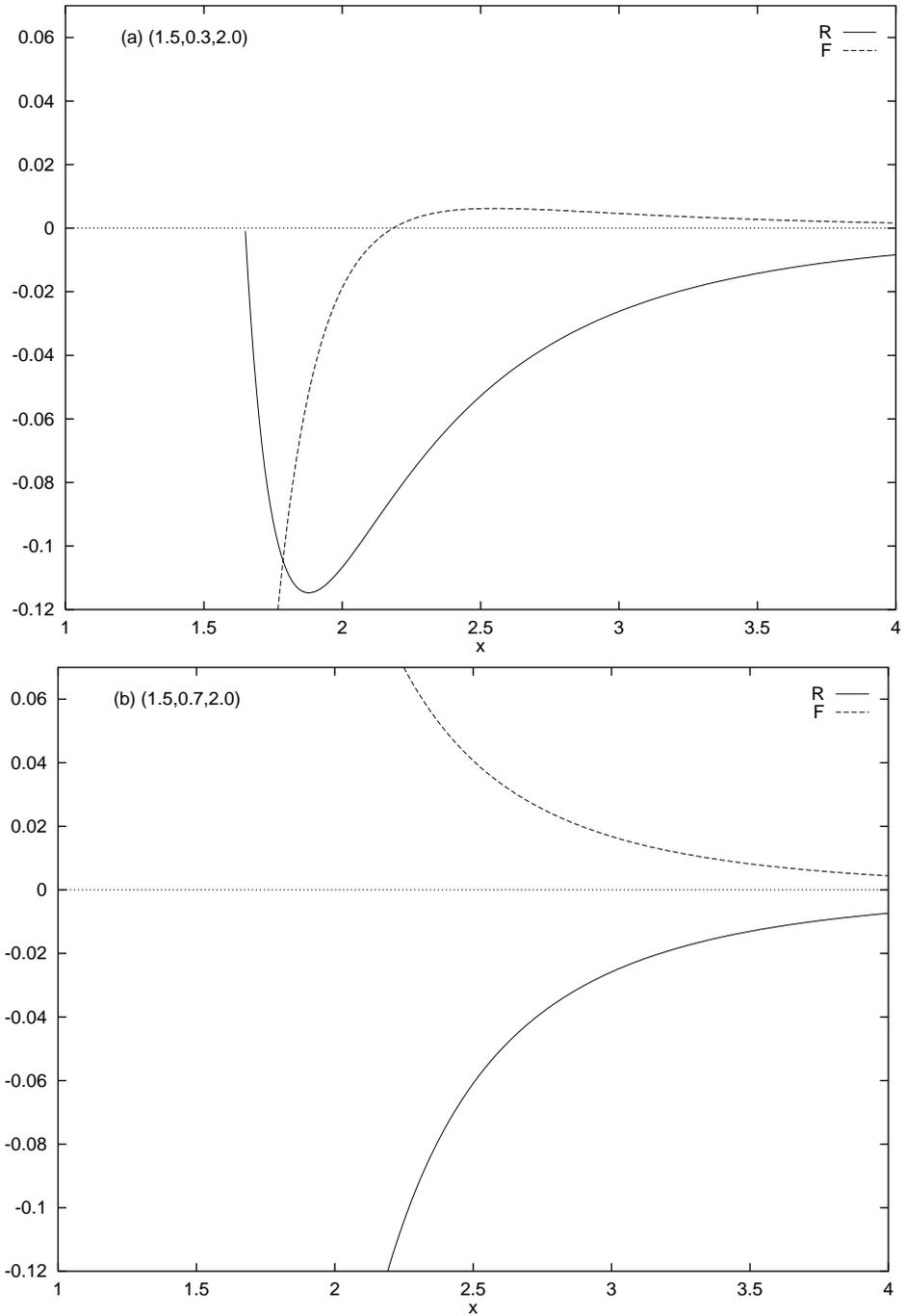

\begin{center}
\hspace*{0.5cm}
\hspace*{0.3cm}
\end{center}
\caption
{The Weyl and Ricci deriving terms, $F$ and $\cal{R}$,
 are shown, where $(\Delta, \delta, h) = (1.5,0.3,2,0)$ in $(a)$ 
and $(1.5,0.7,2.0)$ in $(b)$. 
The vertical and horizontal axes denote, respectively, 
the deriving terms and $x$. }   
\label{F2}
\end{figure}

\end{document}